\documentclass{webofc}
\usepackage[varg]{txfonts}   
\usepackage{braket}
\usepackage{slashed}
\newcommand{\be}{\begin{equation}}
\newcommand{\ee}{\end{equation}}
\newcommand{\bi}{\begin{itemize}}
\newcommand{\ei}{\end{itemize}}
\newcommand{\ba}{\begin{array}}
\newcommand{\ea}{\end{array}}
\newcommand{\bea}{\begin{eqnarray}}
\newcommand{\eea}{\end{eqnarray}}

\DeclareRobustCommand{\vect}[1]{
  \ifcat#1\relax
    \boldsymbol{#1}
  \else
    \mathbf{#1}
  \fi}
\begin{document}
\title{Exclusive $c \to u \gamma$ transitions of $B_c$ meson}
%
%

\author{\firstname{Nicola} \lastname{Losacco}\inst{1,2}\fnsep\thanks{\email{nicola.losacco@ba.infn.it}}
}

\institute{Istituto Nazionale di Fisica Nucleare, Sezione di Bari,  Via Orabona 4, I-70126 Bari, Italy
\and
           Dipartimento Interateneo di Fisica ``M. Merlin'', Universit\`a  e Politecnico di Bari, \\ via Orabona 4, 70126 Bari, Italy
          }

\abstract{%
  We study the rare decays of the $B_c$ meson induced by the flavour changing neutral current $c \to u \gamma$ transition. In the Standard Model they are strongly suppressed by the Glashow-Iliopoulos-Maiani mechanism, therefore they  are sensitive to new physics. The difficulty is to get rid of  long-distance contributions.  We study such effects in radiative  $B_c$  transitions both to $B^*$ and to the axial-vector $B_1^{\prime}$  meson.
}
\maketitle
\section{Introduction}
\label{intro}

The decays induced  by the flavour changing neutral current   $c \to u$  transitions play an important role in the search for  new physics (NP) phenomena  \cite{Gisbert:2020vjx}.
In the Standard Model (SM)  the  relevant weak Hamiltonian involves small Wilson coefficients resulting from the efficient GIM cancellation \cite{Buras:2020xsm}:  the corresponding hadronic amplitudes are highly suppressed for $c \to u \gamma$, $c \to u \ell^+ \ell^-$ and $c \to u \nu \bar \nu$  \cite{deBoer:2017que}. Studies devoted to $B_c^+ \to B^{*+}\, \gamma$ concluded that in this channel the long-distance (LD) contributions do not have the overwhelming size as in the case of $D$, $D_s$ and $\Lambda_c$ decays \cite{Fajfer:1999dq}. We reconsider the issue, extending the analysis to the $B_c^+ \to B^{\prime +}_1 \, \gamma$ mode, with $B^{\prime +}_1$ the lightest axial-vector beauty meson. We find that for $B_c$ decays to spin-1 positive-parity $B_u$ excitations, the LD
contributions are comparable to the $B^{* +}$ case. However, a hadronic suppression in the short-distance (SD) amplitude reduces the role of this channel for searching NP.

\section{Effective weak Hamiltonian}
\label{sec-1}

To describe the $c \to u \gamma$ transition we consider  the effective weak Hamiltonian   \cite{deBoer:2017que}
\be
\mathcal{H}_{eff} = 4 \frac{G_F}{\sqrt{2}} \left[ \sum_{q=d,s}  V_{c q}^* V_{u q} \left( C_1 \mathcal{O}_1^{(q)} + C_2 \mathcal{O}_2^{(q)} \right) + \sum_{i=3}^6  C_1 \mathcal{O}_i 
+ \sum_{i=7}^8  \left( C_i \mathcal{O}_i + C_i ^\prime\mathcal{O}_i^\prime \right) \right]   \label{Hamiltot}
\ee
expressed in terms of  current-current operators $\mathcal{O}_{1,2}^{(q)}$,  QCD penguins operators $\mathcal{O}_{3,\dots 6}$  and   electromagnetic and gluon dipole operators
$\mathcal{O}_{7,8}^{(\prime)}$. The effective coefficient $C_7^{\rm eff}$ including two-loop QCD matrix elements of the operators $O_{3,\dots,6}$, is  in the range \cite{deBoer:2017que}
\be
C_7^{\rm eff} \in [-0.00151-i 0.00556|_s +i 0.00005|_{CKM}, -0.00088-i 0.00327|_s +i 0.00002|_{CKM}], \label{c7}
\ee
where the subscripts indicate the contributions to the  imaginary part  from the strong phases and the phases of the CKM matrix.
$C_7^\prime$ turns out to be small, $C^\prime_7 \sim m_u/m_c$, therefore  the operator $\mathcal{O}^\prime_7$  can be neglected in SM.
Eq.~\eqref{c7}  shows  the GIM suppression in the SM for the dipole $\mathcal{O}_7$, since $C_7 \sim 10^{-3}$. This coefficient can be significantly enhanced in a SM extension (BSM) scenario. General bounds can be established   \cite{Adolph:2018hde}:
\be
|C_7|, |C'_7| \lesssim 0.5.
\ee
The amplitude of the transition $B_c(p) \to A(p^{\prime}, \epsilon) \gamma(q,\lambda)$, where A is a $1^+$ state is given by

\bea
\mathcal{A}(B_c(p) \to A(p^{\prime}, \epsilon) \gamma(q,\lambda)) = \left\{A_{PC} \left[p \cdot q g^{\alpha \beta} - q^{\alpha} p^{\beta} \right] + i A_{PV} \varepsilon^{\alpha \beta \mu \nu} p_{\mu} q_{\nu} \right\} \epsilon_{\alpha}^* \lambda_{\beta}^*.
\label{eq:amplitude}
\eea

\section{Short-distance amplitude}
\label{sec-2}

\begin{figure}[!h]
\begin{center}
\includegraphics[width = 0.4\textwidth]{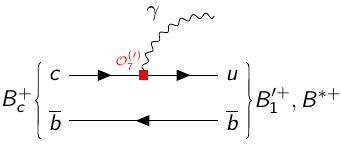}
\includegraphics[width = 0.4\textwidth]{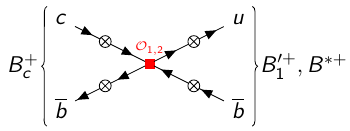}
\caption[Short distance Feynman diagram contribution to rare $c\to u \gamma$ transition.]{\small  Electromagnetic dipole (left) and weak annihilation amplitudes (right) for $B_c \to B^{\prime}_1(B^{*}) \gamma$. The crosses in the WA diagram correspond to the photon emission.}\label{SDFeyn}
\end{center}
\end{figure}

The SD amplitude comes from the $\mathcal{O}^{\prime}_7$ and $\mathcal{O}_{1,2}$ operators Fig. \ref{SDFeyn}.
The weak annihilation (WA) topology is doubly Cabibbo-suppressed, therefore it can be ignored in the evaluation of the SD. 

\noindent 
Focusing on the positive parity channel, the hadronic matrix element is parametrized as
\begin{equation}
\begin{aligned}
\bra{B^{\prime}_1(p',\epsilon)} \bar{u} \sigma_{\mu \nu} c \ket{B_c(p)} &= \frac{\epsilon^* \cdot q}{\left(m_{B_c}+m_{B^{\prime}_1}\right)^2}\left( p_{\mu}p'_{\nu} - p_{\nu}p'_{\mu}\right) T^{\prime}_0(q^2)  \\
&+ \left( p_{\mu}\epsilon^*_{\nu} - p_{\nu}\epsilon^*_{\mu}\right)  T^{\prime}_1(q^2) +  \left( p'_{\mu}\epsilon^*_{\nu} - p'_{\nu}\epsilon^*_{\mu}\right) T^{\prime}_2(q^2), \label{FFT}
\end{aligned}
\end{equation}
\noindent
with $q = p-p^\prime$,  $\lambda$ and $\epsilon$ the photon and $B^{\prime}_1$ polarization vectors. Since the $\bar{u} \sigma_{\mu \nu} \gamma_5 c$ matrix element is obtained using  $\sigma_{\mu \nu} \gamma_5 = -\frac{i}{2} \varepsilon_{\mu \nu \alpha \beta} \sigma^{\alpha \beta}$, with $\varepsilon^{0123}=-1$, the parity-conserving and parity-violating SD amplitudes in Eq. \eqref{eq:amplitude} are given by
 
\be
A^{SD}_{PC} = i \frac{G_F}{ (2 \pi)^{3/2}} m_c \alpha^{1/2} (C^{eff}_7 + C^{\prime}_7) (T_1^{\prime}(0)+T_2^{\prime}(0) )  , \label{SDamplPC}
\ee
\be
A^{SD}_{PV} = - i \frac{G_F}{ (2 \pi)^{3/2}} m_c \alpha^{1/2} (C^{eff}_7 - C^{\prime}_7) (T_1^{\prime}(0)+T_2^{\prime}(0)).
\label{SDamplPV} 
\ee
\noindent
To determine the form factors $T_i$  we use an approach based on the heavy quark spin symmetry \cite{Jenkins:1992nb,Colangelo:1999zn,Colangelo:2022lpy,Colangelo:2022awx}. For a generic Dirac matrix $\Gamma$, the matrix elements of quark currents for negative and positive parity doublets,  invariant under rotations of the heavy quark spins, can be obtained using the trace formalism \cite{Falk:1991nq}:
\be
\bra{B'_1(v,k)} \bar{u} \Gamma c \ket{B_c(v)} = -\text{Tr}\left[ \bar{S} \Omega' \left(v, a_0 k \right) \Gamma H \right],
\label{Trace1} 
\ee
where $S$ and $H$ are the positive and negative parity 
spin doublet fields representation of the states $( B_c, B^*_c )$ and $( B_0, B'_1 )$.
The form factor $\Omega^\prime$ encoding the nonperturbative dynamics is written as \cite{Jenkins:1992nb,Colangelo:1999zn}
\be
\Omega^{\prime} \left(v, a_0 k \right) = \Omega^{\prime}_1\left(v, a_0 k \right)  + \slashed{k} a_0 \Omega^{\prime}_2\left(v, a_0 k \right) .
\ee
The  dimensionful parameter $a_0$ represents the length scale involved in the process: we use the same value for both matrix elements.
Since $m_c \ll m_b$, the $b$ quark remain almost unaffected in the $B_c$  transition, and the velocity of the final meson is nearly unchanged.
Using Eq.~\eqref{Trace1}  we obtain for $B'_1$
\be
\bra{B'_1(v,k; \epsilon)} \bar{u} \sigma_{\mu \nu} c \ket{B_c(v)} = - i \sqrt{ \frac{m_{B_c}}{m_{B'_1}}} \left[ \Omega'_1 \left(k_{\mu} \epsilon^*_{\nu} - k_{\nu} \epsilon^*_{\mu}\right) + 2 a_0 \Omega'_2 m_{B'_1} \epsilon^* \cdot v \left( k_{\mu} v_{\nu} - k_{\nu} v_{\mu} \right) \right]. \label{TME}
\ee
\noindent
The form factors $\Omega_{1,2}^{\prime}$ can be obtained using a determination of matrix elements in a non perturbative approach.
In \cite{Shi:2016gqt} using a covariant light-front approach  the form factors for the
 $B_c \to \left( B_{sJ}, B_{dJ} \right)$ decays in  SM have been determined. We can use these form factors assuming isospin symmetry.
 
The heavy quark spin symmetry  produces relations allowing to determine the tensor form factors. We obtain the finite mass form factors in term of $\Omega_1^{\prime}$ and $a_0 \; \Omega_2^{\prime}$:

\begin{align}
T_0^{\prime}(q^2)&= 2 i {(m_{B_c}+m_{B_1^{\prime}})^2 \sqrt{m_{B_1^{\prime}}} \over m_{B_c}^{3/2}} a_0 \, \Omega_2^{\prime},\\
T_1^{\prime}(q^2)&= - {m_{B_1^{\prime}} \over m_{B_c}} T_2^{\prime}(q^2),\label{relT1T2}\\
T_2^{\prime}(q^2)&= - i \sqrt{{m_{B_c}} \over m_{B_1^{\prime}}} \Omega_1^{\prime}.
\end{align}
Eq. \eqref{relT1T2} together with Eqs. \eqref{SDamplPC}-\eqref{SDamplPV} shows that the SD contribution for the positive-parity final state is suppressed due to a cancellation between two terms.

\section{Long-distance amplitude}
\label{sec-3}

\begin{figure}[!h]
\begin{center}
\includegraphics[width = 0.3\textwidth]{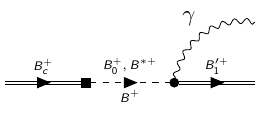}
\includegraphics[width = 0.3\textwidth]{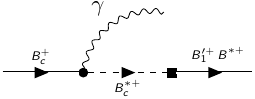}
\includegraphics[width = 0.3\textwidth]{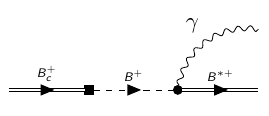} \\
\includegraphics[width = 0.4\textwidth]{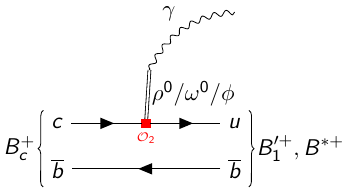}
\caption[Long distance Feynman diagram contributions for $B_c \to  B^{\prime}_1 \gamma$ and $B^{*}$.]{\small LD contribution to $B_c \to B^{\prime}_1 \gamma$ and $B^{*}$. Top panel: Weak annihilation contribution, where the box represent the insertion of a weak operator. Bottom panel:  long-distance pole contribution for the $B_1^{\prime}$ and $B^*$ cases.}\label{LDFeyn}
\end{center}
\end{figure}

LD contributions correspond to processes involving intermediate hadrons. There are two kind of terms. The first one comes from the weak annihilation (WA) amplitude in Fig.~\ref{LDFeyn} (top panel) with the photon radiated by any quark. The second type of contributions are pole terms induced by Eq. \eqref{Hamiltot} with intermediate neutral vector mesons, as in Fig.~\ref{LDFeyn} (bottom panel). For $B_c \to B^{\prime}_1 \gamma$ the intermediate $\rho^0$, $\omega$ and $\phi$ mesons are far from the kinematical range, therefore their contribution is suppressed respect to the case for $B_c \to B^*_u\gamma$.
\noindent 
Let us focusing on $B^{\prime}_1$.

\subsection*{A. WA with intermediate hadrons}
\label{subsec-3.1}
The first weak annihilation diagram in Fig.~\ref{LDFeyn} involves two terms. In the first one the $B_c$ has weak transition to  $B$, $B_0$  and $B^*$, coupled to  $B^{\prime}_1$ and a photon:  
\be
\mathcal{A}^{B_c \to B^{\prime}_1 \gamma} = \mathcal{A}^{B_c \to B_{\text{res}}} \frac{i}{\tilde{p}^2-m_{B_\text{res}}^2} \mathcal{A}^{B_\text{res} \to B^{\prime}_1 \gamma}. \label{LDamp}
\ee 
\noindent
 $B_{\text{res}}$ is one of the off-shell  states $B$, $B_0$ and $B^*$, the  momentum $\tilde p$ is such that $\tilde{p}^2 = m_{B_c}^2$. $\mathcal{A}^{B_c \to B_{\text{res}}}$  can be computed from the Hamiltonian \eqref{Hamiltot} using factorization: 

\be
\mathcal{A}^{B_c \to B_{\text{res}}} = \frac{G_F}{\sqrt{2}} V^*_{cb}V_{ub} a_1  \bra{B_{\text{res}}} \bar{u} \gamma^{\mu}\left( 1 - \gamma_5 \right) b \ket{0} \bra{0} \bar{b} \gamma_{\mu}\left( 1 - \gamma_5 \right) c \ket{B_c} \label{eq:ann}
\ee
\noindent
where $a_1 = C_1 + \frac{C_2}{3}$. The matrix elements in the amplitude involves the decay constants of mesons in the process, $f_{B_c}$,$f_{B_{\text{res}}}$.
The matrix elements of two mesons and the photon can be written in terms of effective couplings $g_{1,2,3}$,
which can be determined by light-cone QCD sum rules (LCSR) in HQET  \cite{Colangelo:2005hv}. Using Eq.~\eqref{LDamp},  the LD contributions  are given by

\begin{align}
\mathcal{A} \left(B_c \to B \to B^{\prime}_1 \gamma \right) &= i \frac{G_F}{\sqrt{2}} V_{cb}^*V_{ub} a_1 f_{B_c}f_{B} \frac{m_{B_c}^2}{m_{B_c}^2-m_{B}^2} e g_1 \left[ \left( \lambda^* \cdot \epsilon^* \right) \left( p' \cdot q \right) - \left( \lambda^* \cdot p' \right) \left( \epsilon^* \cdot q \right) \right] , \\
\mathcal{A} \left(B_c \to B^* \to B^{\prime}_1 \gamma \right) &= 0, \\
\mathcal{A} \left(B_c \to B_0 \to B^{\prime}_1 \gamma \right) &= \frac{G_F}{\sqrt{2}}V_{cb}^*V_{ub} a_1 f_{B_c}f_{B_0}\frac{m_{B_c}^2}{m_{B_c}^2-m_{B_0}^2} e g_3 \varepsilon_{\alpha \beta \sigma \tau} \lambda^{* \alpha} \epsilon^{* \beta} \tilde{p}^{\sigma} q^{\tau}. \label{B0Res}
\end{align}

For the contribution of the second diagram in Fig.~\ref{LDFeyn} the radiative emission from the $B_c$ to the $B_c^*$ is followed by the annihilation to $B^{\prime}_1$. The photon emission amplitude can be parametrized as 
\be
\mathcal{A}^{B_c \to B_c^* \gamma} = \braket{B_c^*(\tilde{p}, \eta) \gamma (q, \lambda)| B_c(p)} = i e g_4 \varepsilon_{\alpha \beta \sigma \tau} \lambda^{* \alpha} \eta^{* \beta} \tilde{p}^{\sigma} q^{\tau},
\ee
with $g_4$ obtained  from the prediction  $\Gamma \left( B_c^* \to B_c \gamma \right) = 33 \, \textrm{eV}$  \cite{Ebert:2002pp}:

\be
g_4^2 = \frac{24 m_{B_c^*}^3 \Gamma \left( B_c^* \to B_c \gamma \right)}{\alpha \left(m_{B_c^*}^2 - m_{B_c}^2 \right)^3}.
\ee
\noindent
For the transition to $B^{\prime}_1$ the amplitude reads:
\be
\mathcal{A}^{B_c^* \to B^{\prime}_1} = \frac{G_F}{\sqrt{2}} V^*_{cb}V_{ub} a_1 \bra{0} \bar{b} \gamma_{\mu}\left( 1 - \gamma_5 \right) c \ket{B_c^*} \bra{B^{\prime}_1} \bar{u} \gamma^{\mu}\left( 1 - \gamma_5 \right) b \ket{0},
\ee
\noindent
with the decay constant $f_{B_c^*}$ and $f_{B^{\prime}_1}$ parametrizing the matrix elements.
The amplitude for the whole process is

\be
\mathcal{A} \left(B_c \to \gamma B_c^*  \to B^{\prime}_1 \right) = \frac{G_F}{\sqrt{2}} V^*_{cb}V_{ub} a_1 f_{B_1} f_{B_c^*} \frac{m_{B_1} m_{B_c^*}}{m_{B_1}^2-m_{B_c^*}^2} e g_4 \varepsilon_{\alpha \beta \sigma \tau} \lambda^{* \alpha} \eta^{* \beta} \tilde{p}^{\sigma} q^{\tau}.
\ee


\subsection*{B. Pole contribution}
\label{subsec-3.2}

Even though the pole contributions should be suppressed, we examine their impact on the $B_1^{\prime}$ channel.

The amplitude for $B_c \to B_1^{\prime} V \to B_1^{\prime} \gamma$,  where $V$ is one of the resonances, reads

\be
\mathcal{A} = \mathcal{A}^{B_c \to B_1^{\prime} V} \frac{i}{q^2-m_V^2+i m_V \Gamma_V} \mathcal{A}^{V \to \gamma}, \label{FullA}
\ee
\noindent
where
\begin{align}
\mathcal{A}^{B_c \to B_1^{\prime} V} &= i \frac{G_F}{\sqrt{2}} a_2 V_{cD}^* V_{uD} m_V f_V \epsilon_V^{* \mu} \braket{B_1^{\prime}(p^\prime,\epsilon)|\bar{u} \gamma_{\mu}(1-\gamma_5) c|B_c(p)}, \label{BcB1primoV}\\
\mathcal{A}^{V \to \gamma} &= -i e {\tilde Q}_D m_V f_V \epsilon_V^{\mu} \lambda_{\mu}^*.
\end{align}
\noindent
$D$ refers to the kind of down quark in the process. For $D=d$ we have ${\tilde Q}_d = {Q_d \over \sqrt{2}}$, for $D=s$ we have ${\tilde Q}_s = Q_s$. 
To evaluate the full amplitude in Eq. \eqref{FullA}, we consider that from gauge invariance the longitudinal helicity amplitude must be discarded \cite{Golowich:1994zr}. This condition is translated into a constraint on the form factor $V_0^{\prime}(0)=0$.

\section{$B_c \to B_u^* \gamma$ }
\label{sec-4}
To test NP in the radiative decay of $B_c$ through the rare transition $c \to u \gamma$, the process $B_c \to B^* \gamma$ is the first one to study. The SD amplitude is given in Eqs.~\eqref{SDamplPC}-\eqref{SDamplPV}, after switching the parity conserving with parity violating expression and with suitable substitutions, $m_{B_1^{\prime}} \to m_{B^*}$ and appropriate tensor form factors parametrizing the matrix element. The LD contributions for this channel are depicted in Fig. \ref{LDFeyn}. Their evaluation follows the same steps as for the LD case of the $B_1^{\prime}$ channel. The LD WA contribution with the $B^+$ resonance is evaluated following \cite{Fajfer:1999dq}. All matrix elements are parametrized as in \cite{Colangelo:2021dnv}, with the form factors for the $B_c \to B_d$ transition constructed using lattice QCD results and heavy quark spin symmetry. Isospin symmetry is also used.

\section{Results}
\label{sec-5}

\begin{figure}[!h]
\begin{center}
\includegraphics[width = 0.5\textwidth]{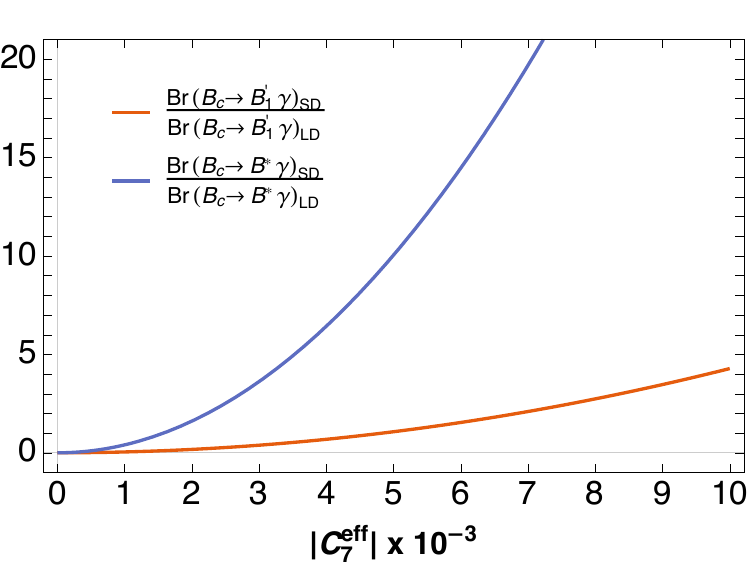}
    \caption[Comparison between the ratio of the branching fractions.]{\small Comparison between the ratio due to SD and LD contributions to the rates of channels with $B^{\prime}_1$ and $B^*$ as final states.} \label{Fig5}   
\end{center}
\end{figure}

In Fig.~\ref{Fig5} we compare the effect of the LD contributions and of the SD one in the $B^*$ and $B_1^{\prime}$ modes.

\noindent
The LD contributions dominate in the small region of $|C^{eff}_7|$, which corresponds to the SM prediction for the Wilson coefficient. If some NP effect increases $C^{eff}_7$,  it would be better observed in the $1^-$ channel.

\section{Conclusions}
\label{sec-6}

Motivated by previous analysis on the rare transition $c \to u \gamma$ in the $B_c$ decays to $B^*$, we focus on the $B_1^{\prime}$ channel to study if it is less polluted by LD contributions.

From a model independent analysis based on heavy quark spin symmetry, we found the relation in Eq. \eqref{relT1T2} between the tensor form factor entering in the SD amplitude for the $B_1^{\prime}$ channel. Their combination in the amplitude produces a suppression of the SD contribution to the process. Therefore, while LD contributions affect more the $B^*$ channel, the hadronic suppression in the $B_1^{\prime}$ case makes it less suitable for accessing NP.

\noindent
\\
{\bf Acknowledgements.}
\begin{acknowledgement}
I thank P. Colangelo and F. De Fazio for collaboration. This study has been carried out within the INFN project (Iniziativa Specifica) SPIF.
The research has been partly funded by the European Union – Next Generation EU through the research Grant No. P2022Z4P4B “SOPHYA - Sustainable Optimised PHYsics Algorithms: fundamental physics to build an advanced society" under the program PRIN 2022 PNRR of the Italian Ministero dell’Università e Ricerca (MUR).
\end{acknowledgement}
\bibliography{proceeding}
%
%
%
%

\end{document}